\documentclass[%
 reprint,
 amsmath,amssymb,
 aps,
 twocolumn,
]{revtex4}

\usepackage{graphicx}
\usepackage{dcolumn}
\usepackage{bm}
\usepackage{physics}
\usepackage{color}
\usepackage{bbm}

\begin{document}

\preprint{APS/123-QED}

\title{Inhomogeneous dynamical mean-field theory of the small polaron problem} 

\author{Kevin-Davis Richler$^1$}
  \author{Simone Fratini$^1$}
  \author{Sergio Ciuchi$^2$}
 \author{Didier Mayou$^1$}
\affiliation{%
$^1$Univ. Grenoble Alpes, Inst NEEL, F-38042 Grenoble, France \\
CNRS, Inst NEEL, F-38042 Grenoble, France\\
$^2$Department of Physical and Chemical Sciences, University of L'Aquila, Via Vetoio, L'Aquila, Italy I-67100\\
Consiglio Nazionale delle Ricerche (CNR-ISC) Via dei Taurini, Rome, Italy I-00185
}

\date{\today}

\begin{abstract}
We present an inhomogeneous dynamical mean field theory (I-DMFT) that is
suitable to investigate electron-lattice interactions in
non-translationally invariant and/or inhomogeneous systems.
The presented approach, whose only assumption is that of a local, site-dependent self-energy,
recovers both the exact solution of an electron for a generic random tight-binding Hamiltonian
in the non-interacting limit and the DMFT solution for the small polaron problem in translationally
invariant systems.
To illustrate its full capabilities, we use I-DMFT
to study the effects of defects embedded on a two-dimensional surface.
The computed  maps of the local density of states reveal Friedel oscillations,
whose periodicity is determined by the polaron mass. This can
be of direct relevance for the interpretation of scanning-tunneling microscopy (STM) experiments
on systems with sizable electron-lattice interactions.
Overall, the easy numerical implementation of the method, yet full self-consistency, allows one to study
problems in real-space that were previously difficult to access.
\end{abstract}

\maketitle

\section{Introduction}{\label{Introduction}
Upon adding an electron to a solid, it will interact
with the dynamical deformation of the periodic lattice. The response of the
medium can be described in terms of a polaron \cite{landau1933ld}, a
quasi-particle which consists of an electron or hole bound together with its
associated lattice polarisation.
While studies on the subject have mostly focused on perfectly periodic, translationally invariant systems, addressing polaronic effects in inhomogeneous media can be of relevance to many real-life situations. Examples include the propagation of electrons (or excitons) from the surface of a sample into the bulk, the propagation of these excitations at the interface
or polaron formation in quantum dots \cite{PhysRevLett.117.116402,hameau1999strong,0953-8984-19-25-255208,bakulin2012role,nourafkan2010metallic,nourafkan2009surface}, to cite a few.
Another important issue which involves inhomogeneous polaron dynamics is related to the influence of local defects and impurities on the propagation of excitations in the bulk.
In general, the many-body nature of the polaronic state does not allow reliable analytical
approaches in the crossover region at intermediate coupling strengths, which is of great interest in
order to quantitatively investigate the conditions of polaron formation.
Therefore, in recent years various numerical methods such as exact
diagonalization \cite{wellein1997polaron,wellein1998self,de1997dynamical},
density matrix renormalization-group (DMRG) \cite{jeckelmann1998density},
diagrammatic Monte Carlo (QMC)
\cite{de1983numerical,kornilovitch1998continuous,kornilovitch1999ground} have been applied to this problem. 
These methods are all quite expensive form a computational point of view and in some cases require analytical continuation to the real frequency axis.\\

To overcome these difficulties, semi-analytical methods based on non-perturbative approximations were also developed in parallel such as dynamical mean field
(DMFT) approaches \cite{freericks1993holstein,ciuchi1997dynamical}, momentum average approximation (MA) \cite{berciu2006green,goodvin2006green} as well as
variational ans\"atze \cite{barivsic2002variational}.
These semi-analytical methods allow a direct real frequency evaluation of the Green function with reduced computational efforts even in three-dimensional cases, but their application to non-translationally invariant systems is far from being straightforward.\\

In disordered materials as well as at interfaces polaron formation occurs in a non-homogeneous environement. It has been pointed out that disorder and more generally inhomogeneity of the media has a positive interplay with polaron formation \cite{ebrahimnejad2012trapping,DiSantePRL2018.118AndrersonHolstein} and such interplay could explain some features of the ARPES studies in oxides at interfaces \cite{NiePRL20152DEG}.
To tackle this problem the inhomogeneous version of the momentum average (IMA) approximation for the single-polaron in inhomogeneous media and coupling with lattice modes \cite{berciu2010holstein} has been introduced. As in the translational invariant case, it can be in principle iterated in a chain of approximations (IMA-$n$) to converge to exact results. However, the computational cost rapidly increases with $n$.\\

The aim of the present work is to provide a general and efficient method for electron-phonon interacting problems in systems lacking translational invariance using the Inhomogeneous Dynamical Mean Field Theory (I-DMFT). 
As in the homogeneous case, I-DMFT will provide an interpolation between the non-interacting case, in which it gives the exact solution of the problem in any disordered and/or patterned geometry, and the strong coupling limit in which it also becomes exact. 
The I-DMFT approximation has been already proven successful in treating strongly correlated spatially inhomogeneous systems including the correlation-driven metal-insulator (Mott) transition at a solid surface \cite{potthoff1999surface,potthoff1999metallic,potthoff1999dynamical}, the spin dynamics in correlated electron systems \cite{lechermann2016electron}, the electronic and transport properties of molecular junctions \cite{jacob2010dynamical}, the effects of electron correlations in Josephson junctions \cite{miller2001microscopic}, the role of lattice defects \cite{delange2016large,backes2016hubbard}, the transport in multilayered inhomogeneous devices \cite{chen2007electronic,zlatic2017thermoelectric} and the correlation of fermions in three-dimensional optical lattices \cite{helmes2008mott}. 
However conventional I-DMFT suffers one major computational limitation since for the calculation of the hybridization function one needs to compute repeatedly the diagonal of the inverse  of a complex matrix, whose dimension equals the number of lattice sites.
Using conventional linear-algebra algorithms, this problem grows cubic with the system size \cite{freericks2016generalized}. 
The formalism that we propose to solve the I-DMFT equations does not require the inverse of a complex matrix. 
Instead the electron-phonon problem is solved on the full lattice under the approximation that the electron-phonon self-energy is local depending   on the frequency of the local phonon only.
Self-consistency equations are than expressed in Hilbert-space such that the recursion technique by Haydock \cite{haydock1980solid} can be used which makes this method immediately generalizable to any lattice geometry and/or disorder distribution.  
Besides, as a result of this self-consistency, the I-DMFT approximation is shown to improve over the existing IMA-1 approximation.
A similar approach has been proposed by \cite{brejnak1995recursion,turchi1997real,julien2001real} to solve the self-consistent equations in the coherent potential approximation (CPA).\\

This paper is organized as follows.
In section \ref{The small polaron problem} we introduce the Holstein
molecular crystal model that we use to study the inhomogeneous polaron problem at zero temperature. 
Section \ref{Chain model representation} briefly introduces Haydock's recursion method, which is used to compute electronic properties of solids without recurring to periodicity or regularity in the structure.
In section \ref{Calculating the Green's function} we discuss the core of the I-DMFT formalism by considering the simplest case of an homogeneous system. 
In section \ref{R_DMFT} we show how to apply the I-DMFT formalism to non-translationally invariant and/or inhomogeneous systems.
Section \ref{Results} is divided into two subsections: A)
we compare our approach with the state-of-the-art numerical solutions of the small polaron problem
B) we apply our method to the study of the local Density of States (LDOS) measured by tunneling and suggest a possible way to measure locally the polaron effective mass.
Finally, section \ref{Summary and Conclusion} gives a brief conclusion and outlook.
\section{Inhomogeneous Holstein model}\label{The small polaron problem}
We consider throughout this paper the Hamiltonian of a single electron interacting with local phonons subject to some
inhomogeneity. We describe this situation using the following generalization of the Holstein model. Its Hamiltonian reads \cite{holstein1959ann}:
\begin{eqnarray} \label{Holstein Hamiltonian} \nonumber
\text{H} = &-& \sum_i\epsilon_i c^{+}_{i}c_{i} - \sum_{<i,j>} t_{i,j}c^{+}_{i}c_{j} \\
       &-&   \sum_{i} g_i c_{i}^{+}c_{i} (a^{+}_{i} + a_{i}) +  \sum_{i} \omega_i a^{+}_{i} a_{i}
\end{eqnarray}
where $c^{+}_i$ and $c_i$ are the creation and destruction operators of
electrons on site i, $a^{+}_i$ and $a_i$ are the creation and destruction
operators of phonons on site i, and the electron's spin index is omitted.
The homogeneous Holstein model is defined by the homogeneous transfer integrals between
nearest neighbors of a $d$-dimensional lattice
$t_{i,j}=t$, the homogeneous phonon frequency $\omega_i=\omega_0$, and the homogeneous strength of the local electron-phonon interaction $g_i=g$.
From these one can construct two independent control parameters \cite{Feinberg1991,SWB-283531347}. The first one is the dimensionless electron-phonon coupling $\lambda=g^2 /D \omega_0$ where $D$ is the
half-bandwidth. The second control parameter is the adiabatic parameter $\gamma = \omega_0 / D$.
While $\lambda$ is the relevant coupling parameter in the adiabatic case ($\gamma <1$), $\alpha^2=g^2/\omega^2_0$ will be the relevant one in the non-adiabatic ($\gamma >1$) case \cite{Feinberg1991,Piovra1998}.
We stress here that, even in the single electron homogeneous case the dressing of the electron by a
coherent multi-phonon cloud, moving coherently with it so as to form a
quasiparticle is difficult to describe within standard perturbative techniques.
Throughout this paper, we will consider explicit inhomogeneity given by the first term in Eq. (1), which describes a site-dependent inhomogeneous onsite energy.
However, the method we devise is applicable to any general Hamiltonian of the type of Eq. (1), with any/all of the microscopic parameters varying in space.
\section{The recursion method \label{Chain model representation}}
For completeness, we recall here a general version of the recursion method \cite{haydock1980solid} which, once adapted to the present problem, can be useful to discuss the self-consistent approximation we use.
Our general problem will be the calculation of the local Green's function of a given state $\ket{\phi_0}$ by constructing an orthonormal basis of states in which the Hamiltonian is tridiagonal. The tridiagonal form of the Hamiltonian allows for a direct visual representation as a semi-linear chain model, which will be key throughout this paper.\\

It is convenient to work in the Krylov subspace of dimension n \cite{gutknecht2007brief}, which is the linear subspace spanned by:
\begin{equation}
\mathcal{K}_{n} (\ket{\phi_0}) = span \left \{ \ket{\phi_0}, \text{H}\ket{\phi_0}, \dotsc , \text{H}^{n-1}\ket{\phi_0} \right \}
\end{equation}
where H is the Hamiltonian and $\ket{\phi_0}$ an initial, normalized, reference state.
An orthogonal basis of the Krylov subspace can be constructed with the Lanczos method \cite{Lanczos1950zz}. This method is an iterative procedure that is capable of constructing the Krylov space via a recurrence relation:
\begin{equation} \label{haydock recursion relations}
\text{H} \ket{\phi_n} = a(n) \ket{\phi_n} + b(n) \ket{\phi_{n+1}} + b(n-1) \ket{\phi_{n-1}}
\end{equation}
with the initial conditions $\ket{\phi_{-1}}=0$, $b(-1)=0$ and where $\ket{\phi_n}$ obey the orthogonality relation $\bra{\phi_n}\ket{\phi_m}=\delta_{n,m}$.  The orthonormal states $|\phi_n \rangle$ are given by
\begin{equation}
\ket{\phi_n} = \sum_{p=0}^{n} c_p \text{H}^{p} \ket{\phi_0}
\end{equation}
where the wave function $\ket{\phi_n}$ will progressively extend away from the initial, local orbital $\ket{\phi_0}$.
The Hamiltonian is tridiagonal in this new basis with diagonal elements $a(0), ... , a(n)$ and  off-diagonal elements $b(0), ... , b(n)$:
\begin{equation}
 \text{H}_{ \mathcal{K}_{n}} = \left[ \begin{matrix}
a(0) & b(0) & 0 & 0  & \cdots & \\
b(0) & a(1) & b(1) & 0 & 0 & \cdots   \\
 0 &  b(1) & a(2)  & b(2) & 0  & \cdots \\
\vdots  & \ddots & \ddots & \ddots & \ddots & \ddots  \end{matrix}  \right] \, .
 \end{equation}
Once written in the $\ket{\phi_n}$ basis the Hamiltonian is thus equivalent to a
semi-infinite one dimensional tight-binding model with nearest neighbor hopping. 
This chain model is determined by its energy-independent hopping integrals $b(n)$ and energy levels $a(n)$ and can be described by the pictorial representation shown in Fig. \ref{semi_linear_chain} a).
Since this method does not rely on any symmetries of the original Hamiltonian it is also applicable where lattice periodicity and/or homogeneity are lost.
The chain model representation is a conceptually attractive, always feasible and numerically exact method for the calculation of the Green's function, that allows computing directly the 
local Green's function defined as
\begin{equation}
G_{00}(z) = \bra{\phi_0} (z-\text{H})^{-1} \ket{\phi_0} 
\end{equation}
without the explicit knowledge of the eigenstates and eigenvectors, in terms of its coefficients $a(n),b(n)$ \cite{haydock1980solid}.
This is achieved by evaluating the continued fraction expansion:
\begin{equation} \label{CF} 
G_{00}(z) = \cfrac{1}{ z - a(0) -
		\cfrac{b(0)^2}{ z - a(1) -
		\cfrac{b(1)^2}{ \, \dots }}} \, .
\end{equation}
In actual calculations it is customary to set $z=E + i\eta$ with $\eta$ an infinitesimally small number, leading to a Lorentzian broadening of the spectral features.
Since Eq. (\ref{CF}) can be written in the form $1/(z- a(0) - \Sigma(z))$, the self-energy $\Sigma(z)$ can also be obtained from the same continued fraction expansion.
The knowledge of the recursion coefficients, therefore provides an alternative way of reconstructing the local Green's function and thus the local density of states, which is more time-efficient than the full diagonalization of the tridiagonal and/or original Hamiltonian. If not specifically
needed and unless otherwise stated we shall use Eq.
(\ref{CF}) throughout this work.\\

In addition to computing efficiently the Green's function, one can use the chain model to represent exactly the effect of an energy dependent self-energy. 
For one lattice site, we can depict the effect of a local self-energy $\Sigma(z)$ as in Fig. \ref{semi_linear_chain} b) by attaching a semi-linear chain to this site (proof in Appendix \ref{Self_energy_appendix}). 
Of course this semi-linear chain must have the proper coefficients a(n), b(n) that correspond to $\Sigma(z)$.
This equivalence is a central concept throughout this paper. 

\section{Calculation of the Green's function: Homogeneous Case}\label{Calculating the Green's function}
Our aim is the calculation of the local retarded Green's function which is defined as
\begin{equation}
      \label{Green function}
G_{i,i}(z) = -i \int_{0}^{\infty} \text{dt}  \bra{VC}T c_i(t) c^{+}_i(0)\ket{VC} e^{izt}
\end{equation}
where $\ket{VC}$ represents the vacuum state for phonons and electrons.
 \begin{figure}[!t]
  \centering
    \includegraphics{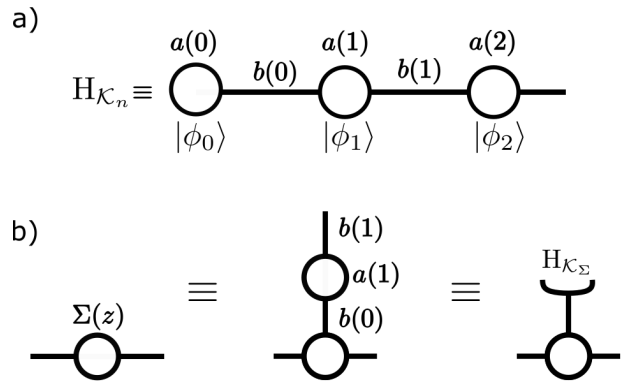}
    \caption{a) Tight-binding representation of the tridiagonal Hamiltonian $\text{H}_{ \mathcal{K}_{n}}$. The recursion method maps the Hamiltonian onto a semi-infinite chain of atoms, where the a(n), b(n) and  $\ket{\phi_n}$ are the energy levels, hopping integrals and states of the orthonormal Lanczos basis of the n-th orbital, respectively. b) Compact pictorial representation which is used throughout this paper, where $\text{H}_{ \mathcal{K}_{\Sigma}}$ labels the tight-binding representation of $\Sigma$.} 
    \label{semi_linear_chain}
 \end{figure}
We here consider the local Green function for the DMFT solution of the homogeneous model Eq. (\ref{Holstein Hamiltonian}) when all $\epsilon_i=0$.
In DMFT we can write the local Green's function as
\begin{eqnarray}
      \label{local Green DMFT}
G(z)^{-1} = z -\Delta(z) - \Sigma(z)
\end{eqnarray}
where $\Delta(z)$ is the hybridization function. 
Since the system is translationally invariant $\Delta$ does not depend on the site index $i$ \cite{georges1996dynamical}.
DMFT relies on a self-consistency relation, i.e. $\Delta(z)$ and $\Sigma(z)$ are mutually dependent functionals
\begin{equation} \label{self_consistentcy_trans} 
\Delta(z) = F_{\Delta}[\Sigma(z)] \quad \text{and} \quad \Sigma(z) = F_{\Sigma}[\Delta(z)].
\end{equation}
The very reason for which they are functionals and not functions depends on retardation effects induced by electron-phonon interaction. 
In standard DMFT procedures a self-consistent solution of Eq. \ref{self_consistentcy_trans} is obtained iteratively at each complex energy $z$.
In this paper, we propose an alternative approach of solving Eq. \ref{self_consistentcy_trans} which is based on Haydock's recursion scheme applied to suitably defined Hamiltonians $\text{H}_{\Sigma}$, $\text{H}_{\Delta}$. This is shown in the following.\\

Because of the analytical properties of $\Sigma(z)$ and $\Delta(z)$, one can expand both by a continuous fraction expansion:
\begin{equation} \label{hybridization_cf}
 \Delta(z) = \cfrac{b_{\Delta}(0)^2}{ z - a_{\Delta}(1)
          - \cfrac{b_{\Delta}^2(1)}{ z -a_{\Delta}(2) - \dots  } } 
\end{equation}
and
\begin{equation} \label{self_energy_cf}
 \Sigma(z) = \cfrac{ b_{\Sigma}(0)^2}{ z - a_{\Sigma}(1)
          - \cfrac{b_{\Sigma}^2(1)}{ z -a_{\Sigma}(2) - \dots } } \, .
\end{equation}
where by this definition both are independent of the onsite energy of the orbital.
The idea of the present formalism is to calculate the continuous fraction coefficients $b_{\Delta}(n)$, $a_{\Delta}(n)$, $b_{\Sigma}(n)$ and $a_{\Sigma}(n)$. 
Once a sufficiently large number of the continued fraction coefficients is determined, both functions are computed using the in section \ref{Chain model representation} present method.\\

In the following we derive a representation of $\Delta$, $\Sigma$ by tight-binding Hamiltonians $\text{H}_{\Delta}$, $\text{H}_{\Sigma}$ which by construction fulfill the 
Eq. \ref{self_consistentcy_trans}.
This is achieved by replacing self-energies and hybridization functions by their equivalent semi-linear chain representation. 
The proof of replacing $\Sigma(z)$, $\Delta(z)$ by semi-linear chains is shown in Appendix \ref{Self_energy_appendix}.
Using Haydock's recursion scheme can then be used to compute the continued fraction expansion of $\Sigma$ and $\Delta$.
\begin{figure}[!t]
 \centering
   \includegraphics{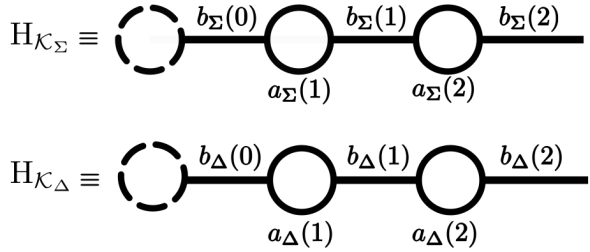}
   \caption{Tight-binding representation of the tridiagonal Hamiltonians $\text{H}_{\mathcal{K}_{\Sigma}}$ and $\text{H}_{\mathcal{K}_{\Delta}}$ as energy independent chains. The self-energy (hybridization function) on the first site (dashed) of the chain equals $\Sigma$ ($\Delta$). By this definition both are independent on the on-site energy of the first site.}
   \label{hybridization_selfenergy}
\end{figure}
 \begin{figure}[!b]
  \centering
    \includegraphics{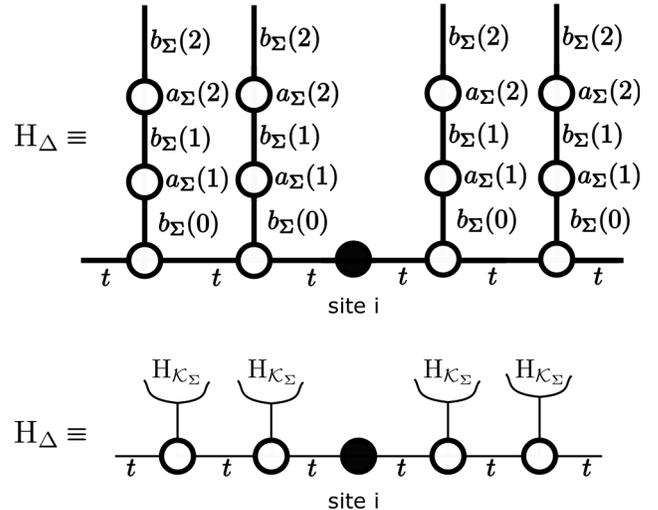}
    \caption{ Upper: Tight-binding representation of the Hamiltonian $\text{H}_{\Delta}$ (non-tridiagonal) which gives rise to the hybridization function at site i. The black marked site labels the initial state of the recursion method, where by construction the system attached to it fulfills $\Delta(z) = F_{\Delta}[\Sigma(z)]$. The self-energy is independent of the onsite energy of the black marked site. Lower: Compact pictorial representation of $\text{H}_{\Delta}$. This representation will be used throughout this paper for  $\text{H}_{\Delta}$.
    }
    \label{spread_wave_function}
 \end{figure}
The DMFT approximation amounts to consider a site-dependent but local self-energy  $\Sigma(z)$ on each lattice site. The tight-binding representation of the hybridization function at site i can then be obtained by suppressing the self-energy at site i and replacing each $\Sigma(z)$ by its equivalent tight-binding representation as a semi-linear chain where the chain representation of $\Sigma(z)$ is depicted in Fig. \ref{hybridization_selfenergy}.
The corresponding tight-binding representation of the Hamiltonian $\text{H}_{\Delta}$ describing the hybridization function is given graphically (in $\text{d}=1$) in Fig. \ref{spread_wave_function}, where the black marked site labels the initial, local wave-function of the recursion method.
The system attached to this site is equivalent with $\text{H}_{\mathcal{K}_{\Delta}}$.
A generalization to higher dimension and/or geometries changes the representation of $\text{H}_{\Delta}$ only. 
This is precisely what makes the proposed formalism straightforward to apply to any lattice geometry, device shape or distribution of disorder.
Using the explicit expressions of $\Sigma(z)$ \cite{ciuchi1997dynamical}
\begin{equation}
      \label{DePolarone Sigma}
 \Sigma(z) = \cfrac{g^2}{ z - \omega_0 - \Delta(z-\omega_0)
          - \cfrac{2g^2}{ ... } }
\end{equation}
allows for a graphical representation of the tight-binding Hamiltonian $\text{H}_{\Sigma}$ (describing the self-energy) by substituting the hybridization function $\Delta(z)$ by its tight-binding representation as a semi-linear chain where the chain representation of $\Delta(z)$ is depicted in Fig. \ref{hybridization_selfenergy}.
The tight-binding Hamiltonian $\text{H}_{\Sigma}$ describing the self-energy is shown in Fig. \ref{local_green_function} where the black marked site labels the initial state of the recursion method. The system attached to this site is equivalent with $\text{H}_{\mathcal{K}_{\Sigma}}$.
It is evident, that $\Sigma(z)$ depends on a discrete set of $\Delta(z-n\omega_0)$.\\ 

Since the coefficients in Fig. \ref{spread_wave_function} and \ref{local_green_function} are initially unknown, we are now going to explicitly demonstrate a procedure to compute recursively starting from an initial state localized on a single lattice site. This is possible due to the progressive extension of the wave-function away from the initial, local orbital.  
The recursion on $\text{H}_\Sigma$, $\text{H}_\Delta$ must be done in parallel as continued fraction coefficients calculated at step n from $\text{H}_\Sigma$ ($\text{H}_\Delta$) are needed to pursue the recursion procedure at step n+1 in $\text{H}_\Delta$ ($\text{H}_\Sigma$).
One key point is that the self-consistency condition required in standard DMFT formulations, is automatically enforced by this construction with no extra step needed.\\

 \begin{figure}[!t]
 \centering
\includegraphics{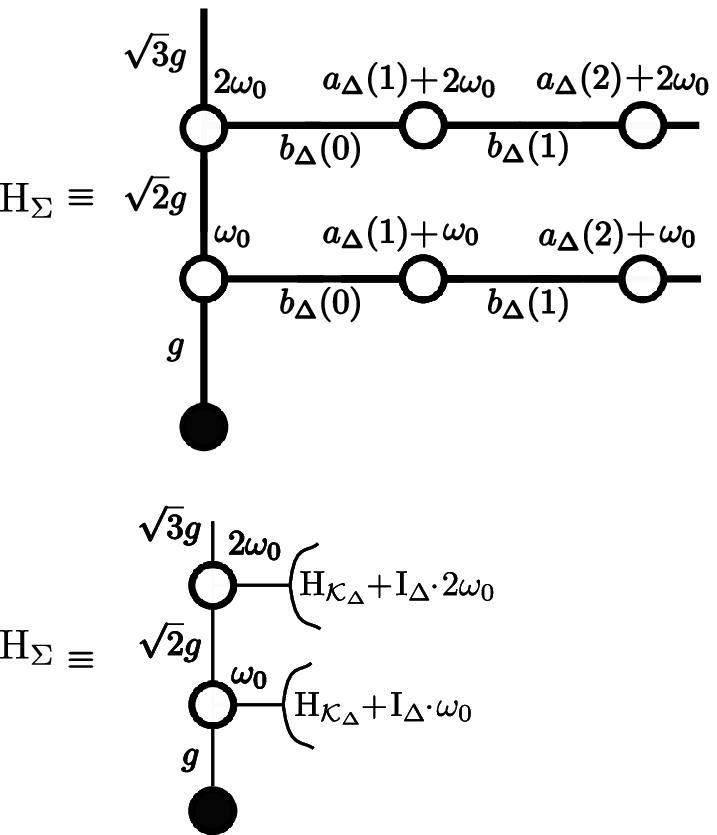}
  \caption{ Upper: tight-bindfing representation of the Hamiltonian $\text{H}_{\Sigma}$ (non-tridiagonal) which gives rise to the self-energy at site i. 
The black marked site labels the initial wave-function of the recursion method, where by construction the system attached to it fulfills $\Sigma(z) = F_{\Sigma}[\Delta(z)]$. The self-energy is independent of the onsite energy of the black marked site.
 Lower: Compact pictorial representation of $\text{H}_{\Sigma}$ where $\text{I}_{\Delta}$ is the unity matrix in the subspace $\mathcal{K}_{\Delta}$. This representation will be used throughout this paper for $\text{H}_\Sigma$. 
  }
   \label{local_green_function}
\end{figure}
The coefficients are determined starting with an initial recursion state which corresponds to an orbital centered on a given site (full black circle in the figures) and is denoted as $\ket{\phi_0}$, 
$\ket{\psi_0}$ respectively, where $\text{H}_{\Sigma}$ acts on $\ket{\phi_0}$, while $\text{H}_{\Delta}$ acts on $\ket{\psi_0}$ only.
Upon iterating the recursion procedure,
the vectors $\ket{\phi_n}$ ($\ket{\psi_n}$) extend away from the initial site, acquiring
non-zero components on sites up to a distance n. 
Thus one needs to know coefficients up to n-1 only to compute the wave-functions $\ket{\phi_n}$ ($\ket{\psi_n}$). This is schematically shown in Fig.
\ref{recursion_scheme} a).
The values of $a_{\Sigma}(n)$, $b_{\Sigma}(n)$ can thus
be determined recursively provided that we know all the coefficients of the series
$a_{\Delta}(0)$, $b_{\Delta}(0)$, $a_{\Delta}(1)$, $b_{\Delta}(1)$, $\dots$ up
to $a_{\Delta}(n-1)$, $b_{\Delta}(n-1)$. 
Based on the above, it is straightforward to show by a precise analysis of the above recursion scheme, that one can generate an explicit
expression for the coefficients $a_{\Sigma}$, $b_{\Sigma}$ in terms of the coefficients
$a_{\Delta}$, $b_{\Delta}$:
\begin{equation} \label{coeff_sigma}
a_{\Sigma}(n) = a_{\Sigma}[a_{\Delta}(n-1),b_{\Delta}(n-1), \dots ] \, ,
\end{equation}
\vspace{-0.7cm}
\begin{equation} \label{coeff_sigma_2}
b_{\Sigma}(n) = b_{\Sigma}[a_{\Delta}(n-1),b_{\Delta}(n-1), \dots ] \, ,
\end{equation}
and vice versa:
\begin{equation} \label{coeff_delta}
a_{\Delta}(n) = a_{\Delta}[a_{\Sigma}(n-1),b_{\Sigma}(n-1), \dots ] \, ,
\end{equation}
\vspace{-0.7cm}
\begin{equation} \label{coeff_delta_2}
b_{\Delta}(n) = b_{\Delta}[a_{\Sigma}(n-1),b_{\Sigma}(n-1), \dots ] \, ,
\end{equation}
These relations have the equivalent meaning as Eq. (\ref{self_consistentcy_trans}) and are assuring self-consistency. 
A detailed computation of the first two sets of recursion coefficients is given in Appendix \ref{Recursion_appendix}.
A representation of the recursion scheme is summarized in Fig. \ref{recursion_scheme} b). 
Note that the in Fig. \ref{recursion_scheme} b) presented recursion scheme does not represent any self-consistency loop, which can be found in standard DMFT approaches \cite{georges1996dynamical}.
This peculiarity reduces computational complexity significantly.
In Appendix \ref{DMFT_gauge} we gauge our numerical results with the DMFT results derived in \cite{ciuchi1997dynamical}, which considers a Bethe lattice of infinite coordination number. In all tested cases, we obtain a remarkable agreement of both methods. 

\begin{figure}[!t]
  \centering
    \includegraphics{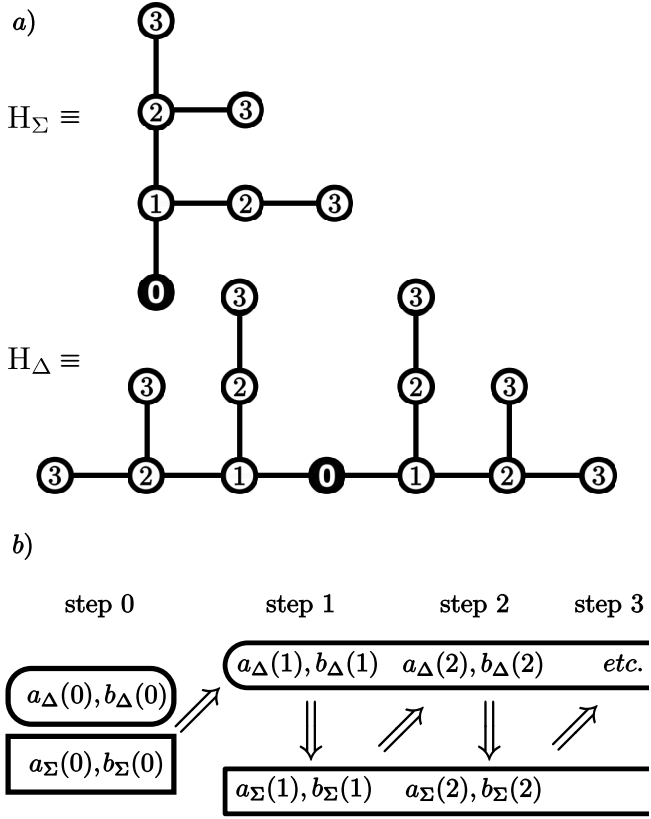}
    \caption{a) The most outward extension of the recursion vector for the first 3 steps for the Hamiltonians $\text{H}_{\Sigma}$ and $\text{H}_{\Delta}$. b) In step n of the recursion scheme, one computes the recursion coefficients of $\text{H}_{\mathcal{K}_{\Delta}}$ using previously calculated coefficients. Knowing these  coefficients allows to purse the computation of a new pair of coefficients of $\text{H}_{\mathcal{K}_{\Sigma}}$.}
    \label{recursion_scheme}
 \end{figure}
\section{Inhomogeneous systems: the I-DMFT method} \label{Non-translational invariant system} \label{R_DMFT}

Given that  the derivation of the previous section was explicitly performed in real space,
it can be straightforwardly generalized to systems lacking translational invariance once the approximation of local self-energy has been considered. Therefore, we assume that the self-energy is local but site-dependent, i.e. $\Sigma_{i,j}(z) \to \Sigma_{i}(z) \delta_{i,j}$.
Within the local self-energy approximation the local Green's function reads:
\begin{equation}
G_{ii}(z)^{-1} = z - \epsilon_i - \Delta_i(z) - \Sigma_i(z) \, .
\end{equation}
For illustrative purposes we consider here the one-dimensional inhomogeneous Holstein Hamiltonian presented in Eq. \ref{Holstein Hamiltonian}.
The computation of the Green function $G_{ii}$ is for one given realization of disorder and therefore $\Sigma_i$ and $\Delta_i$ are disorder dependent quantities.
In the following we will show, that in the absence of translational invariance one has to solve N local impurity problems by performing 2N recursion in parallel exchanging coefficients after each recursion step where this is obviously well suited for a practical implementation via parallel computing.\\ 

As in the previous section, one can define the tight-binding representation of the self-energy
$\Sigma_i(z)$ and the hybridization function $\Delta_i(z)$ for every site.
In Fig. \ref{local_green_function_invariant} we show the self-energy and the
corresponding hybridization function of the impurity site and its neighboring site
located on its left.\\ 

Starting from the I-DMFT Eqs. presented in section (\ref{R_DMFT}) one can easily see that the
self energy on a specific site $\Sigma_i(z)$ depends on the hybridization
function $\Delta_i(z)$. Formally, one can write down the following
functional relation:
\begin{eqnarray} \label{non_equivalent_self_energy}
\Sigma_i(z) = F_{\Sigma_i}[ \Delta_{i}(z) ] \, .
\end{eqnarray}
Following the same lines as in Sec. \ref{Calculating the Green's function} one
can express these functions as continuous fraction expansions and show that, for each site $i$, the
recursion coefficients of the self-energy are given by a set of recursion
coefficients of the hybridization function analogous to Eqs. (\ref{coeff_sigma}), (\ref{coeff_sigma_2}):
\begin{equation}  \label{non_equivalent_self_energy_1}
a_{\Sigma_i}(n) = a_{\Sigma_i}(n)[ a_{\Delta_i}(n-1),b_{ \Delta_i}(n-1), \dots] \, ,
\end{equation}
\vspace{-0.7cm}
\begin{equation}   \label{non_equivalent_self_energy_2}
b_{\Sigma_i}(n) = b_{\Sigma_i}(n)[ a_{\Delta_i}(n-1),b_{\Delta_i}(n-1), \dots] \, .
\end{equation}
In the case of non-equivalent sites
the different hybridization functions depend on a set of self-energies.
Formally, one can write down the following functional relation:
\begin{eqnarray} \label{non_equivalent_hybridization}
\Delta_i(z) = F_{\Delta_i}[ \{ \Sigma_{j}(z) \} ]_{\substack{ j\neq i }}
\end{eqnarray}
which includes all self-energies except the self-energy $\Sigma_i(z)$ on site i.
These functions can be expressed as continuous fraction expansions
where the recursion coefficients of the self-energy are given by the set
of recursion coefficients of the hybridization function:
\begin{equation} \label{coeff_delta_inhomo_a}
a_{\Delta_i}(n) = a_{\Delta_i}(n)[ \{ a_{\Sigma_j}(n-1),b_{\Sigma_j}(n-1) \} ]_{\substack{ j\neq i }} \, ,
\end{equation}
\vspace{-0.7cm}
\begin{equation} \label{coeff_delta_inhomo_b}
b_{\Delta_i}(n) = b_{\Delta_i}(n)[ \{ a_{\Sigma_j}(n-1),b_{\Sigma_j}(n-1) \} ]_{\substack{ j\neq i }} \, .
\end{equation}
In contrast to the previous set of relations (\ref{non_equivalent_self_energy_1}),(\ref{non_equivalent_self_energy_2}) these equations now do couple different sites, via their corresponding hybridization functions. 
Again through the hierarchical construction of recursion coefficients self-consistency is assured and thus no self-consistency loop is needed.\\

 \begin{figure}[!t]
  \centering
    \includegraphics{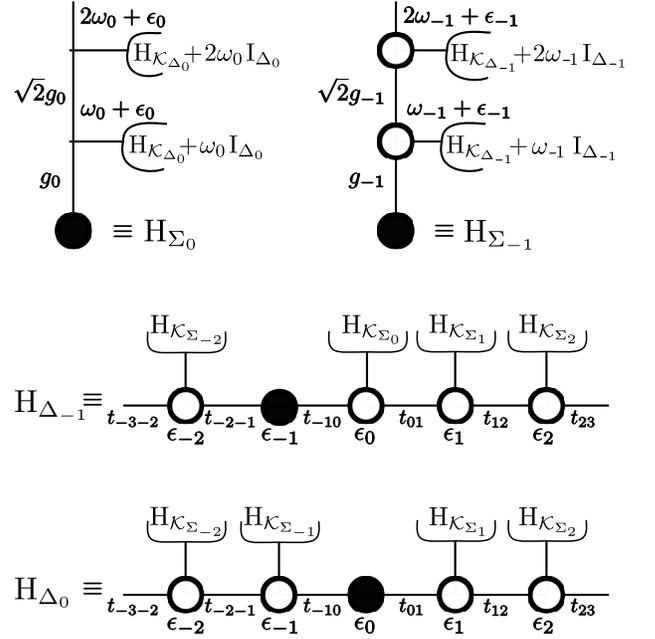}
    \caption{Tight-binding representation of the Hamiltonian $\text{H}_{\Sigma_{0}}$, $\text{H}_{\Sigma_{-1}}$, $\text{H}_{\Delta_{0}}$ and $\text{H}_{\Delta_{-1}}$ with $\text{I}_{\Delta_{n}}$ being the unity matrix in the subspace $\mathcal{K}_{\Delta_n}$. $\text{H}_{\Delta_{0}}$ and $\text{H}_{\Delta_{-1}}$ is a two-dimensional representations of the 1d lattice problem including the phonon space.}
    \label{local_green_function_invariant}
 \end{figure}
 
Finally, let us comment on general numerical aspects. 
The Holstein model has been studied previously by means of exact or direct Lanczos diagonalization preserving the full Hilbert space \cite{wellein1998self,wellein1997polaron,de1997dynamical}. 
The full Hilbert space has dimensions N that grow exponentially with the system-size being given by the number of phonons that are kept on each site to the power of the number of lattice sites.
Since exact (Lanczos) diagonalization techniques require $\mathcal{O}(N^3)$ ($\mathcal{O}(N^2)$) floating point operations (flops) to diagonalize the Hamiltonian matrix, actual calculations are restricted to extremely small  systems (few lattice sites only) due to computer memory and CPU limitations, which is clearly not suitable for the description of inhomogeneous systems. 
The aim of I-DMFT is to reduce the full Hilbert space, such that calculations can be performed at large system sizes, allowing to fully address the relevant spatial variations of the physical properties in the presence of inhomogeneities, but still affording a good description of the physical processes of interest. 
By greatly reducing the size of the Hilbert space, which now grows linearly with the system size (it is given by the product of the number of lattice sites times the number of phonon per site), and the fact that
Haydock's recursion requires $\mathcal{O}(N)$ flops only, the present method can easily handle homogeneous, inhomogeneous systems of up to $10^4$, $10^3$ lattice sites respectively in a reasonable time (hours).

\section{Results}\label{Results}
\subsection{Local impurity on one dimensional lattices}
As a first example, we consider the Holstein Hamiltonian for a single defect at the center of the lattice $i=0$, i.e. $\epsilon_i = -\abs{U} \, \delta_{i,0}$ and compare our numerical results with the inhomogeneous momentum average approximation (IMA-1) and the diagrammatic Monte Carlo (DMC) results derived by \cite{berciu2010holstein}. 
In order to deal with a single or a cluster of impurities we further improve the efficiency of the algorithm by assuming that only a part of the total size of the system (N) is affected by inhomogenetiy. We thus define a number $\text{N}_\text{i}<\text{N}$ of non-equivalent lattice sites centered around the impurity of the cluster. The number is chosen in such a way that results are independent of $\text{N}_\text{i}$. 
In particular, we have found that for a single attractive impurity a cluster of 10 non-equivalent sites (impurity + cluster of sites surrounding the impurity) is sufficient in 1d, which is in agreement with the observation by \cite{berciu2010holstein}. 
This assumption optimizes the computational complexity and costs significantly and thus will be used throughout this paper to treat clusters of finite disorder.\\

For the Holstein model the computational requirements can be further optimized. 
The number of possible phonon configurations is infinite, but can be restricted to a finite, sufficiently large number in actual calculations.
In particular, we have found, that choosing M (maximum number of phonon excitations per site) approximately two (one) orders of magnitude higher than $\alpha^2$ is sufficient to captures the polaron formation and thus to compute the local density of states (ground-state properties), which is in agreement with the observation by \cite{Feinberg1991}. 
This is achieved by cutting the vertical chain in the chain model representation of $\text{H}_\Sigma$ (see Fig. \ref{local_green_function}) after M sites.
This assumption reduces the computer memory usage significantly and will be used throughout this paper.\\ 

 \begin{figure}[!t]
  \centering
    \includegraphics{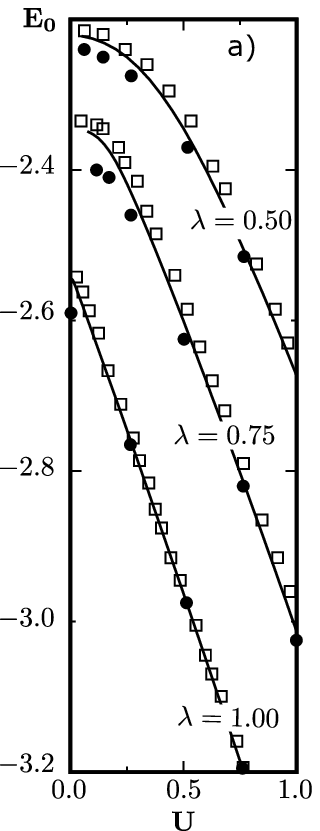} \hspace{0.1cm}
        \includegraphics{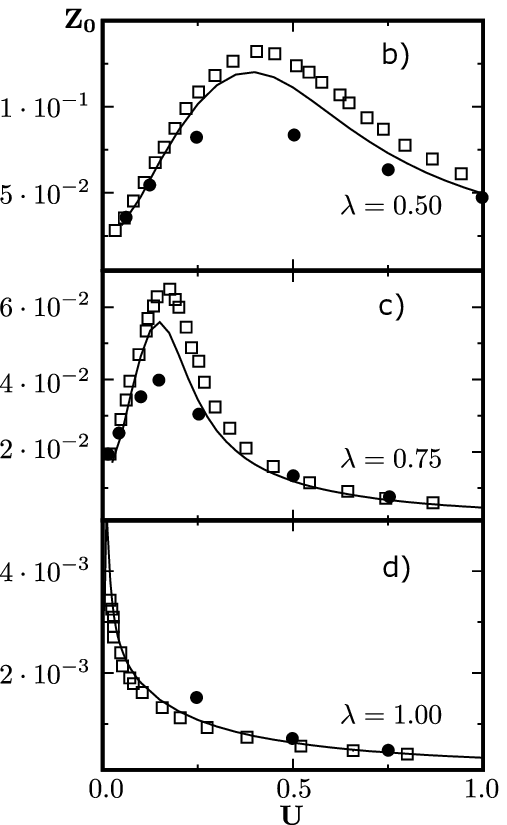}
    \caption{ We present the ground-state energy a) and the quasiparticle weight b)-d) as a function of the onsite potential U for $\gamma=0.2$ and $\lambda=0.5,0.75,1.0$ (black line). Circles (squares) represent the DMC (IMA-1) results derived in \cite{berciu2010holstein}.}
    \label{IMA}
 \end{figure}

In addition, I-DMFT needs a large system in order to have a sufficiently dense sampling of states (recursion coefficients), since finite size effects naturally introduce a limitation on the value of $\eta$, where this kind of finite size effects is inherent to the recursion method and is present even in the non-interacting limit. 
We have found, that approximately $10^3$ ($10^{2}$) recursion coefficients are sufficient in order to compute the local density of states (ground-state properties), while assuring a sufficient spectral resolution of $10^{-2}$ (limiting the error to $10^{-3}$).
The dimension of the total system, i.e. the number of total lattice sites, the number of phonons per site and the cluster size of non-equivalent lattice sites is given explicitly for every calculation.\\

In this and the following, all energies are expressed in units of the total bandwidth.  
In Fig. \ref{IMA} is depicted the ground-state energy $E_0$ and the quasiparticle weight $Z_0$ at the impurity site as a function of the impurity potential $U$ in the adiabatic regime.
We present our I-DMFT results (N $\approx 1,000$, M $\approx 50$, $\text{N}_{\text{i}}$ $\approx 50$) for the worst-case scenario.
I-DMFT, being a local approximation, is less accurate in one space dimension and in the limit of low-phonon frequency at a small intermediate value of the coupling. 
Large polarons are expected to be pinned by the impurity in this case.
Although there are quantitative differences between I-DMFT and DMC, on can see in Fig. \ref{IMA} that the qualitative agreement between I-DMFT and DMC is good and as expected, it improves as $U$ increases.
Noticeably the ground state energy is always lower (i.e. more accurate) than that of IMA-1 for any value of $U$. 
This trend is also observed in the quasiparticle spectral weight when compared with IMA-1 and DMC i.e. I-DMFT is always in between IMA-1 and DMC no matter if DMC predicts a smaller or larger quasiparticle spectral weight.
\subsection{Local impurity on two dimensional lattices}
 \begin{figure}[!b]
  \centering
    \includegraphics[scale=0.750]{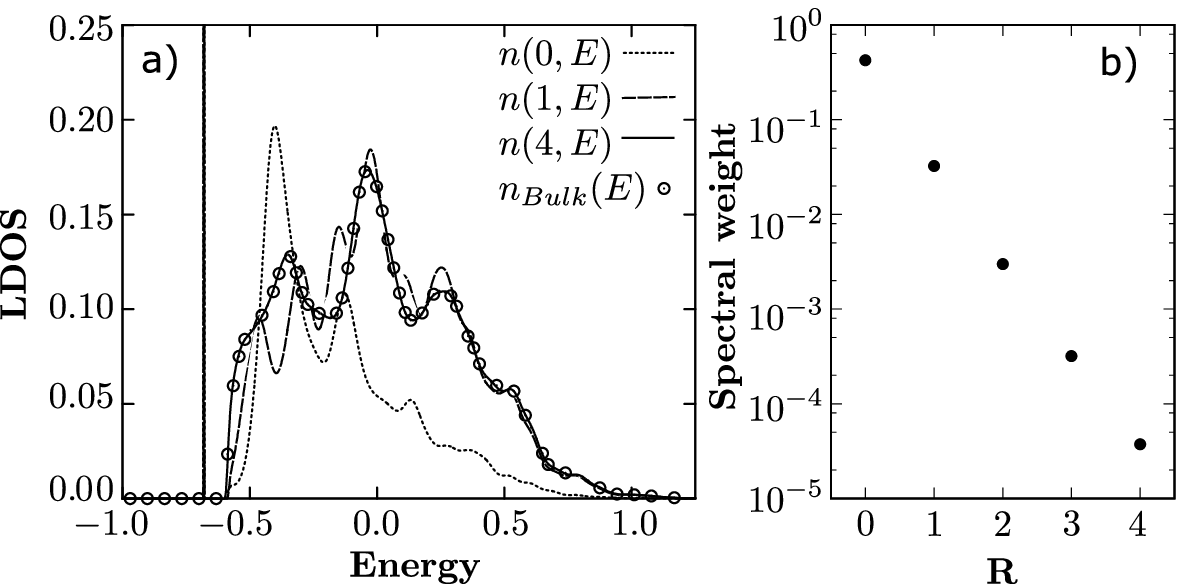}

     \hspace{0.4cm}
     
  \includegraphics[scale=1.0]{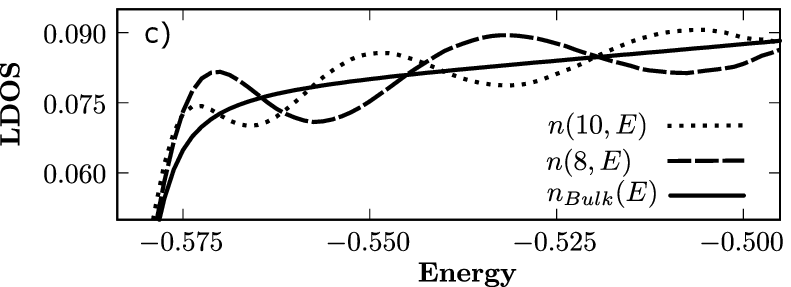} \vspace{0.15cm}
\caption{a) LDOS $n(R,E)$ for several distance along direction $(1,0)$ on a square lattice compared with that in absence of the impurity which represent the bulk value $n_{Bulk}(E)$.
b) Spectral weight of the localized state.
c) Friedel oscillations in the LDOS $n(R,E)$ as a function of energy. All calculations are done for $\gamma=0.5$, $\lambda=0.4$ and $U=0.38$.
}
    \label{one_impurity_deviations}
 \end{figure}

We here consider the case of an impurity in a two dimensional square lattice, i.e. a single defect located at the origin of the lattice $i=0$.
In Fig. \ref{one_impurity_deviations} we plot the spectral density at site $i$ (N $\approx 800$, M $\approx 80$, $\text{N}_{\text{i}}$ $\approx 9\times9$)
for several values of the distance from the impurity $R=R_i-R_0$
\begin{equation}
n(R,E)=-\frac{1}{\pi}\Im G_{i,i}(z)
\end{equation}
compared with that obtained in absence of the impurity $n_{Bulk}(E)$ where $z$ is defined by $E + i\eta$ and $\eta$ an infinitesimal small number. 
The spectral resolution in our calculations are limited by a) the finite sampling of states (recursion coefficients), which naturally introduces a limitation on $\eta$, and b) the continuous fraction expansions which enforces Lorentzian broadening. 
This leads to an underestimation of the critical localization strength where bound states are found below the continuum.
Since I-DMFT reduces greatly the full Hilbert space, 
we have diagonalized the chain model representation of every Green's function $G_{ii}(z)$ using Lanczos method and have computed the LDOS by 
applying a Gaussian broadening in order to obtain a continuous spectrum.
Applying Lanczos method increases the spectral resolution (Gaussian instead of a Lorentzian broadening) and gives direct access to the spectral weight and eigenvalues by only increasing the numerical cost slightly. 
This allows to determine the critical disorder strength more accurately and will be used throughout this paper to decide whether bound-states appear below the energy continuum.
We can see, that a bound state appears below the energy-continuum at approximately $\text{E}_{\text{loc}}\approx-0.68$ redistributing the continuum spectrum. The spectral weight associated with the bound state is shown for greater clarity in Fig. \ref{one_impurity_deviations} b). Its exponential decay gives the localization length of the bound state.\\

 \begin{figure}[!b]
  \centering
        \includegraphics[scale=1.2]{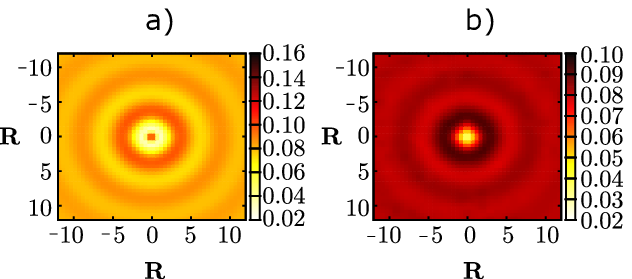} 

                \vspace{0.5cm}

        \includegraphics[scale=1.]{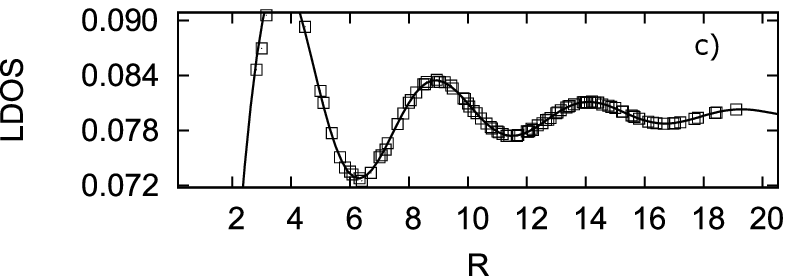}         
                \includegraphics[scale=1.]{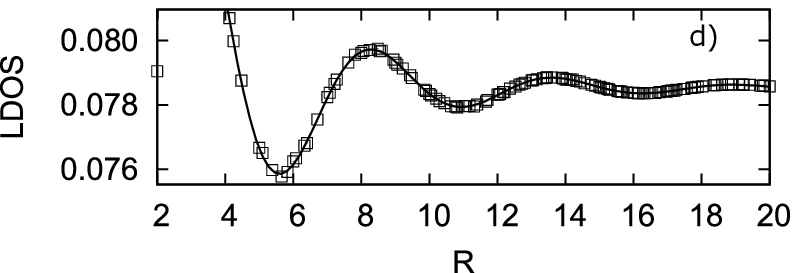} 
                   \caption{a) and b): Maps of the spectral density of states $n(R,E)$. c) and d) Friedel oscillations in the LDOS as a function of distance R (crosses) to the Impurity, together with a fit using Eq. (\ref{friedel_oscillation}) (solid line), where values of the LDOS at identical R have been averaged. a),c) have been computed for $\lambda=0.4$, $\gamma=0.5$, $U=0.38$, $E=-0.54$, while b),d) have  been computed for $\lambda=1.0$, $\gamma=0.5$, $U=0.38$, $E=-0.69$.}
    \label{friedel}
 \end{figure}
The presence of a defect causes Friedel oscillations in the density of states due to the loss of translational invariance \cite{friedel1952xiv}. In Fig. \ref{friedel} the calculated Friedel oscillations in the local density of states are shown as a function of distance (N $\approx 500$, M $\approx 5$,  $\text{N}_{\text{i}}$ $\approx 20\times20$). We focus on the regime of elastic electron-phonon scattering, i.e. where $E_0 < E < E_0 + \omega_0$ fluctuations of the density of states around $n_{Bulk}(E)$ can be approximated far away from the defect as
\begin{equation} \label{friedel_oscillation}
n(R,E) = n_0(E) + \delta n(E)  \cdot \frac{\text{cos}(2 k R  + \chi )}{R}
\end{equation}
where $R$ is the distance from the defect,  $\chi$ is a phase shift,  $k=\sqrt{(E_0-E + i\eta)/Z_0}$ and $\delta n = \epsilon_0/2\pi^2 k$. From Friedel's oscillations one can identify the quasiparticle which scatters from the impurity as the polaron.
Using Eq. (\ref{friedel_oscillation}) one can indeed compute the effective mass ($m^{*}/m=1/Z_0$) of the quasiparticle from the interference pattern by fitting Eq. (\ref{friedel_oscillation}) to the computed numerical data which we illustrate for $\gamma=0.5$ and for two different values of the bare coupling constant, i.e. $\lambda=0.4,1.0$.
The fits are shown in \ref{one_impurity_deviations} (c)-(d).
We find that the effective mass is given approximately by $m^{*}/m \approx 1.25$, $1.83$, respectively. 
This is in good agreement with the values calculated from the self-energy $\Sigma_{Bulk}(z)$ of the bulk $m^{*}/m = 1.22$, $1.87$ considering that the error is of the order $\eta = 5\times10^{-2}$ in both calculations.
The errors in determining the effective mass trough Friedel's oscillations are essentially due to the vanishing of their amplitude as a function of distance to values less than one percent thus reaching the accuracy limit in our numerical procedure.
On the other hand, a large number of coefficients is also needed in order to decrease the damping factor that we include in our calculation due to finite values of $\eta$ which are of the order $5\times10^{-2}$.
\section{Summary and Conclusion \label{Summary and Conclusion}}
We have presented I-DMFT, a powerful method to address electron-phonon interaction in systems lacking translational invariance. The method is based on a local self-energy approximation, which interpolates between the vanishing and the infinite electron-phonon coupling. In particular, it gives the exact solution for a disordered system recovering (when present) Anderson localization. Overall, we have found that the presented I-DMFT formalism is quantitatively accurate and easy to parallelize.
I-DMFT is efficient to study electron-phonon interaction within the Holstein Hamiltonian for arbitrary geometries and disorder configurations.
In particular, I-DMFT allows one to study samples with a large number of non-equivalent sites where one can consider disorder in the on-site energy, the hopping, the electron-phonon coupling constant and the phonon frequency.
We have here presented the case study of isolated impurities, benchmarked our results with results derived in \cite{berciu2010holstein}.
We have found that I-DMFT is more accurate than IMA-1 and gives quantitative accurate results when compared to the approximation free DMC calculations.
As an application of our method, we computed LDOS maps for a single defect on a square lattice, as conventional methods are not suitable due to enormous computational costs.
From an analysis of Friedel oscillation patterns, we demonstrate that it is possible to extract the polaron mass from the periodicity of the oscillation.
Finally, the presented formalism can be easily extended to study the effect of chemical disorder, electron-phonon coupling to several phonon modes and finite temperature. Therefore one can easily combine several extensions in order to study more realistic and complex models. 
\begin{acknowledgments}
We thank D. Feinberg and P. Qu\'emerais for the stimulating discussions and comments.
The author acknowledges the LANEF framework (ANR-10-LABX-51-01) for its support with mutualized infrastructure.
\end{acknowledgments}
\appendix
\section{Equivalence of chains and self-energies \label{Self_energy_appendix}}
We show, that the in Fig. \ref{semi_linear_chain} attached semi-linear chain is equivalent to a self-energy $\Sigma(z)$.
We define the projection operators $P$ and $Q$, such that $P$ projects onto the subspace of the atom, while $Q$ projects onto the subspace of all states of the semi-linear chain.
Using $P+Q=1$, $P^2=P$, $Q^2=Q$ and $PQ=QP=0$ one finds
\begin{equation} \label{PGP}
P\frac{1}{z-\text{H}}P = \cfrac{P}{z - P\text{H}P - P\text{H}Q \cfrac{1}{z-Q\text{H}Q} Q\text{H}P} \, ,
\end{equation}
where we have defined $\Sigma(z)$ as
\begin{equation} 
\Sigma(z) = \bra{\phi_o} P \text{H} Q \frac{1}{z- Q \text{H} Q} Q \text{H} P \ket{\phi_0} \, .
\end{equation}
with $\ket{\phi_0}$ denoting the local orbital at the atom. 
Thus, coupling the atom to a semi-linear chain is equivalent to adding a self-energy $\Sigma(z)$ to it (see Fig. \ref{semi_linear_chain}). 
This is used throughout this paper to replace $\Sigma(z)$, $\Delta(z)$ by their tight-binding representation as a semi-linear chain (see Fig. \ref{hybridization_selfenergy}). 
Of course these semi-linear chains must have proper coefficients that correspond to $\Sigma(z)$, $\Delta(z)$.

\section{Computation of recursion coefficients  \label{Recursion_appendix}}
To further illustrate the recursion scheme, we show in Fig. \ref{appendix_detail_comp} the chain representation of $\text{H}_{\Sigma}$, $\text{H}_{\Delta}$ and compute explicitly the first two sets of recursion coefficients for the one dimensional homogeneous Holstein problem presented in section \ref{Calculating the Green's function}. 
Starting the recursion procedure from $\ket{\psi_0}=\ket{0,0}$ and using Eq. (\ref{haydock recursion relations}) one finds
\begin{eqnarray} \label{A1}
\text{H}_{\Delta} \ket{0,0} &=& t \ket{1,0} + t \ket{-1,0} \nonumber \\
  &=& a_{\Delta}(0) \ket{\psi_0} + b_{\Delta}(0) \ket{\psi_1} 
\end{eqnarray} 
where states in the energy independent chain representation are labeled $\ket{x,n}$ with x being the lattice coordinate and n being the phonon number (see Fig. \ref{appendix_detail_comp}).
Projecting on Eq. \ref{A1} with $\ket{\psi_1}$, $\ket{\psi_0}$ respectively one derives $a_{\Delta}(0)=0$, $b_{\Delta}(0)=\sqrt{2}t$ and the new wave function
\begin{equation}
 \ket{\psi_1} = \cfrac{ \ket{1,0} + \ket{-1,0}}{\sqrt{2}} \, .
\end{equation}
Further starting from $\ket{\phi_0}=\ket{0,0}$ one finds
\begin{eqnarray} \label{A2}
\text{H}_{\Sigma} \ket{0,0} = g \ket{0,1} = a_{\Sigma}(0) \ket{\phi_0} + b_{\Sigma}(0) \ket{\phi_1} \, .
\end{eqnarray}
Projecting on Eq. \ref{A2}) with $\ket{\phi_1}$, $\ket{\phi_0}$ respectively one derives $a_{\Sigma}(0)=0$, $b_{\Sigma}(0)=g$ and the new wave function $\ket{\phi_1} = \ket{0,1}$.
In the second recursion step one derives
\begin{eqnarray} \label{A4}
\text{H}_{\Delta} \ket{\psi_1} &=& t \ket{1,0} + t \ket{-1,0}  \\
&=& a_{\Delta}(1) \ket{\psi_1} + b_{\Delta}(1) \ket{\psi_2} + b_{\Delta}(0)\ket{\psi_0} \, , \nonumber \\
\text{H}_{\Sigma} \ket{\phi_1} &=& \sqrt{2}g \ket{0,2} + \omega_0 \ket{0,1} + b_{\Delta}(0) \ket{1,1}  \\
 &=& a_{\Sigma}(1) \ket{\phi_1} + b_{\Sigma}(1) \ket{\phi_2}  + b_{\Sigma}(0)\ket{\phi_0} \, , \nonumber
\end{eqnarray}
The new set of recursion coefficients are given by $a_{\Delta}(1)=0$, $b_{\Delta}(1)=\sqrt{t^2 + b_\Sigma(0)^2}$, $a_{\Sigma}(1)=\omega_0$, $b_{\Sigma}(1)=\sqrt{2g^2+b_\Delta(0)^2}$ and the new wave functions are
\begin{eqnarray}
b_{\Delta}(1) \ket{\psi_2} &=& \frac{t}{\sqrt{2}} \ket{-2,0} + \frac{b_{\Sigma}(0)}{\sqrt{2}} \ket{1,1}  \nonumber  \\ 
		  &+& \frac{t}{\sqrt{2}} \ket{2,0}  + \frac{b_{\Sigma}(0)}{\sqrt{2}} \ket{-1,1} \, ,  \\
b_{\Sigma}(1) \ket{\phi_2} &=& b_{\Delta}(0) \ket{1,0} + \sqrt{2}g \ket{0,1} \, . 
\end{eqnarray}
 \begin{figure}[!t]
  \centering
    \includegraphics[scale=1.0]{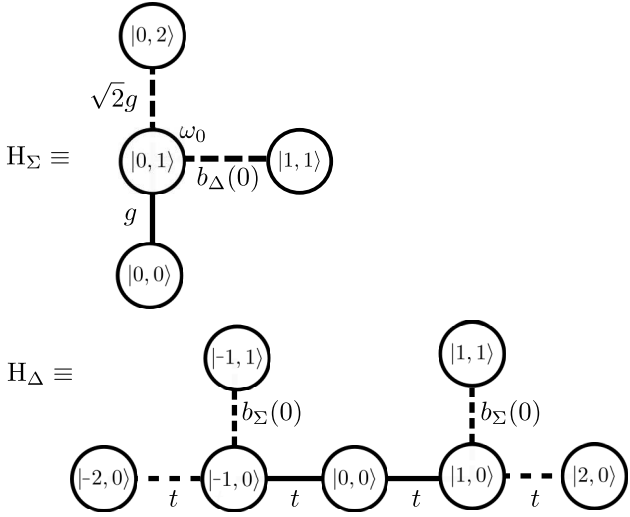}
    \caption{Tight-binding representation of the Hamiltonians $\text{H}_{\Sigma}$ and $\text{H}_{\Delta}$. $\text{H}_{\Delta_{0}}$ and $\text{H}_{\Delta_{-1}}$ is a two-dimensional representations of the 1d lattice problem including the phonon space. Extension of the recursion vector for the first two recursion steps are presented by solid and dashed lines. }
    \label{appendix_detail_comp}
 \end{figure}

First one shall notice, the recursion on $\text{H}_{\Sigma}$ and $\text{H}_{\Delta}$ must be done in parallel, since the coefficients calculated at step 0 from $\text{H}_{\Sigma}$ ($\text{H}_{\Delta}$) are needed to continue the recursion procedure  on $\text{H}_{\Delta}$ ($\text{H}_{\Sigma}$) at step 1. Second, the DMFT approximation amounts to considering a site-dependent but local self-energy $\Sigma$ on each lattice site and thus a generalization to higher dimension and/or geometries changes the representation of $\text{H}_{\Delta}$ only.
\section{Comparison with the DMFT results of Ref. \cite{ciuchi1997dynamical} \label{DMFT_gauge}}
 \begin{figure}[!b]
  \centering
    \includegraphics[scale=0.5639]{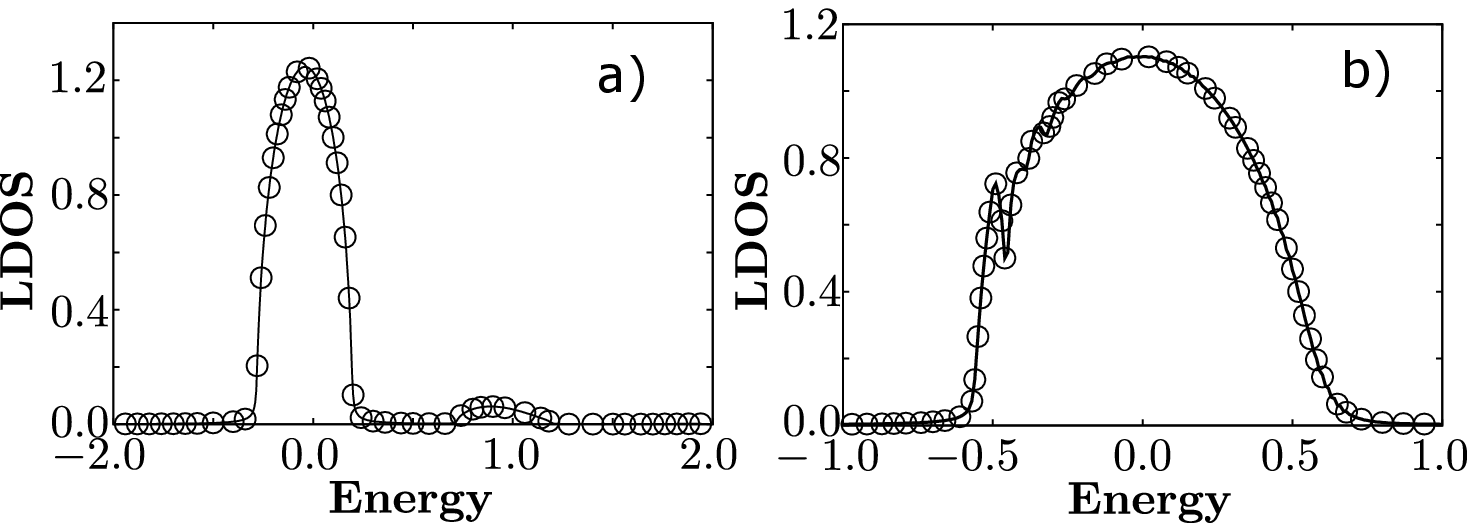}  
                \includegraphics[scale=0.5825]{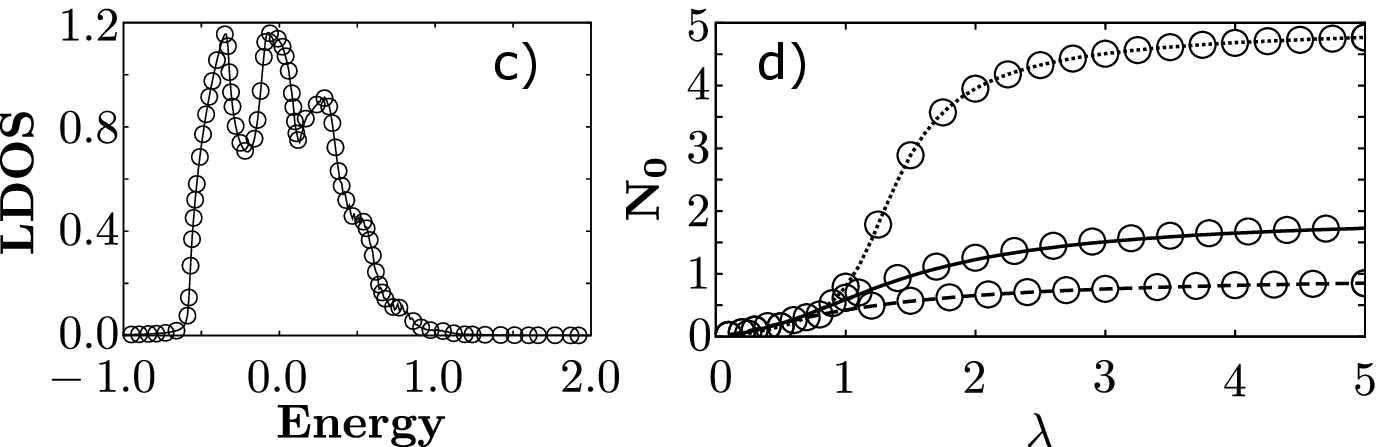}
    \includegraphics[scale=0.65]{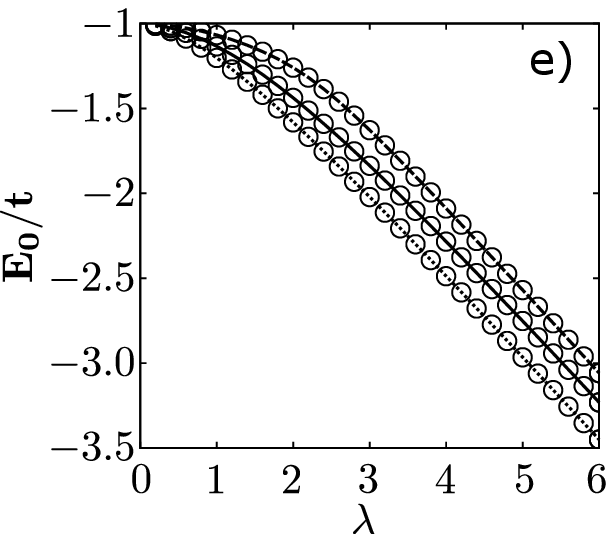} 
    \includegraphics[scale=0.65]{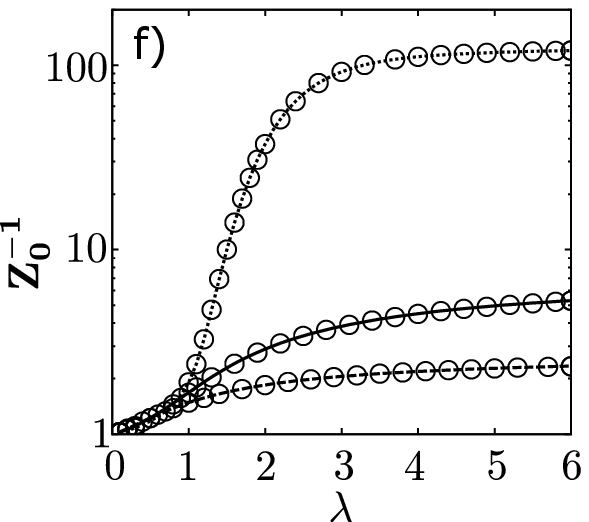}
    \caption{We present the spectral density (solid line) in the large a), low b) and intermediate c) phonon frequency regime. a) is computed for $\lambda=0.08$, $\gamma=2.0$, b) $\lambda=0.7$, $\gamma=0.125$ and c)  $\lambda=0.4$, $\gamma=0.5$. In d) we present as a function of $\lambda$ the average number of phonons, e) ground-state energy and f) quasiparticle weight for $\alpha^2=1,2,5$ (solid, doted, dashed). Throughout a)-f) circles represent the results derived in \cite{ciuchi1997dynamical}.}
    \label{spectral density bethe lattice}
 \end{figure}
 
We here gauge our numerical results with the DMFT results derived in \cite{ciuchi1997dynamical}, which considers a Bethe lattice of infinite coordination number.
Using the Lanczos method, a real lattice can be mapped into a 1d semi-chain, which is used as the underlying lattice with electron hopping parameter b(n) and onsite energy a(n). 
Due to translational symmetry, the information about the lattice, and thus the hybridization to the impurity is completely preserved in this step.
We begin our comparison by discussing the LDOS. In Fig. \ref{spectral density bethe lattice} we present the spectral density for the low, intermediate and large phonon frequency regime  (N $\approx 10000$, M $\approx 100$). As can be seen, the newly developed method gives excellent agreement in the low, intermediate and large phonon frequency regime as both results overlap.
Further, in Fig. \ref{spectral density bethe lattice} a comparison of the ground-state energy, quasiparticle weight and the average number of phonons as a function of $\lambda$ is shown (N $\approx 800$, M $\approx 30$).
As can be seen, the newly developed method gives excellent agreement for all quantities.
One shall notice, that the accuracy was also tested for other values of $\lambda$, $\gamma$ and $\alpha$.
However, in all tested cases, we obtain a remarkable agreement. 

\clearpage
\nocite{}
\bibliography{apssamp-sergio}
\end{document}